\documentclass[intlimits,twoside,a4paper]{article}

\usepackage[cp1251]{inputenc}

\usepackage{amsmath,  amsfonts, amssymb}
\usepackage{graphicx}
\usepackage{bm}
\usepackage{caption}
\usepackage{subcaption}
\usepackage{tikz}
\usepackage{geometry}
\usepackage[eqsecnum]{cmpj3}

\issue{2021}{24}{1}{13302}
\doinumber{10.5488/CMP.24.13302}
\title[Spectral density of interacting pairs in low-dimensional finite lattices]%
{Spectral density of interacting pairs in low-dimensional finite lattices%
}
\author[T. Chattaraj]{T. Chattaraj}
\address{
The University of British Coulambia (UBC), Vancouver, Canada
}

\date{Received July 12, 2020, in final form August 29, 2020}
\begin{document}

\maketitle

\begin{abstract}
The spectral density of bound pairs in ideal 1D, 2D and Bethe lattices is computed for weak and strong interactions. The computations are performed with Green's functions by an efficient recursion method in real space. For the range of interaction strengths within which bound states are predominantly single pairs, the spectral profiles guide to the energy bandwidths where the bound pairs can be maximized.
\end{abstract}

\section{Introduction}
For the Hubbard model \cite{hubb} in lattices, formation of composite objects can be observed for both attractive and repulsive interactions \cite{wink}. A doublon is such a composite object formed from two fermions. In Auger spectroscopy, certain lines and lineshapes in the spectra can be explained within the formalism of formation of bound pairs \cite{sawa, lens,cini}. In general, pairs formed from interacting particles show the  properties different from constituent individual particles. For strongly interacting particles, their lifetimes are generally high \cite{stro} and depend on interaction strengths. These strongly correlated pairs are now realized and observed in optical lattices within broad range of tuneable interaction strengths \cite{prei}. Recently, the pairing phenomena have been also observed in optical lattices  for spin excitations \cite{fuku} which are relevant for excited states near the ground ferromagnetic state. The bound states that arise in lattices from onsite and nearest neighbour interactions are also related to solitons \cite{klei, mill}. 

Two interacting particles show nonclassical correlations depending on the statistics of the particles~\cite{qin}.  The interaction changes the effective statistics between particles. However, a clear mathematical account on this is yet expected to emerge. The constraints of the two-particle density matrices are also not known \cite{mcwe}. One can study the correlations from quantum walk experiments and simulations which are fundamental building blocks of many-body systems. The dynamics of particles in lattices changes significantly depending on the binding. The quantum walk of correlated particles has promising applications in quantum technologies \cite{chil, child, perr}. 

A general understanding on the mechanism of superconductivity is also still unresolved. The Hubbard model with onsite and nearest neighbour interactions is mostly used for explaining the phenomena. Pairing of fermions with attractive interaction is known to cause the mechanism \cite{bard, micn}. In two-dimensional Hubbard model, the super-exchange in quasi two-dimensional oxides and nearest neighbour coulomb interaction between polarons in organics \cite{maju} is known to be responsible for the phenomena. It is suggestive that an unified approach to the mechanisms may have relations to the two-particle bands with weak and strong interactions within the range where two-particle bound pairs have a higher density than the bound states with a larger number of particles.  

The effects of magnetic fields on single particles or excitations in two-dimensional lattices are well known from Hofstadter's approach \cite{hofs}. The density of states for two particles for various interaction strengths had also been computed before \cite{shep}. In optical lattices, the external magnetic fields can be synthesized with periodic modulation of the lattice potentials \cite{lin, dali, stru}. This article specifically looks at the density of bound pairs for both zero and non-zero magnetic fields at different interaction strengths. In the context of condensed matter, the effective fields can also be realized through doping with isotropic magnetic impurities. 

The solution for the two-particle states are known exactly for strong interaction cases which are described with continuum and bound states \cite{petr}. Within many particle environment, there can be several bound states and interplay of these bound states with the continuum states are highly complex. These many-particle bound states can appear at different energy windows with different densities. Green's functions provide a direct approach to study these states at different interaction strengths. Recursion methods \cite{recu} are known for efficient computation of these Green's functions. Several other methods have also been applied recently for such computations, e.g., exact diagonalization with the method of Lanczos \cite{lanc}, DMRG \cite{scho} and so on. In this article, the Green's functions are computed using Berciu's approach \cite{berc} which is best described in the original article. A brief description is given in the method section for the reader. With Berciu's method, the Green's functions can be computed efficiently for the cases of zero and non-zero external magnetic fields as well as for disordered systems~\cite{tirt}. The bound pairs can be found in general graph architectures where the quantum transport may prove relevant to quantum technologies. The Bethe lattice is one such graph for which the density of bound pairs at different interaction strengths is considered in this article. The Bethe lattice which is important for its mathematical form, also has presence in cores of several photosynthetic complexes.

\section{Method}
In this article the density of a bound pair of two interacting spinless particles is studied. The particles are taken as hardcore bosons in 1D and 2D lattices and softcore bosons in Bethe lattice. The case of hardcore bosons is similar to two spinless fermions. In 1D and 2D lattices, the particles get bound with nearest neighbour interactions. For the models with zero nearest neighbour interaction and non-zero onsite interaction, the same bound pairs appear for spin paired particles. The simplified model Hamiltonian is that of extended Hubbard model $\mathcal{H}_{e\text{H}}$ with the hopping limited to the nearest neihbour sites on the lattices. 
  \begin{equation}\label{ham}
\mathcal{H}_{e\text{H}} = \sum_{m} \epsilon_{m} a_{m}^\dagger a_m  + \sum_{<mn>} t_{|m-n|} a_{m}^\dagger a_n + \sum_m U a_{m}^\dagger a_{m}^\dagger a_m a_m + \sum_{<mn>} V_{mn} a_{m}^\dagger a_{n}^\dagger a_n a_m.
 \end{equation} 
The onsite interaction term is taken as infinite in 1D and 2D cases. The nearest neighbour interaction term is neglected for the Bethe lattice. Onsite energies are scaled to zero for regular lattice systems.

For a Hamiltonian $\mathcal{H}$, the Fourier transformed Green's function is defined as
     \begin{equation}\label{gf}
              G(\omega) = \frac{1}{\omega - \mathcal{H}},
       \end{equation} 
    where $\omega = E + \ri\eta$ is a complex number with $\eta$ a very small positive real number. In real space, the two particle Green's function $ G_2(m,n,\omega) = \langle m n | G(\omega) | m' n' \rangle $
is the propagator for two particles from sites $m'$, $n'$ to sites $m$, $n$ in a lattice, where $| m n \rangle = c_m^\dagger c_n^\dagger |0\rangle$ and the initial site indices $m'$, $n'$ are omitted for brevity. The bound pair spectral density can be calculated for the bound state from the Green's function $G_2(m', n', \omega)$ in real space  with initial distance $|m' - n'|=0$ and $1$ for softcore and hardcore bosons,  respectively 
   \begin{equation}\label{spec}
     A (m', n', E) = \frac{-1}{\piup}  \mbox{Im}[G_2(m', n', E + \ri\eta)].
    \end{equation}
The total density of states (DOS) is obtained with summing over densities of all the states
   \begin{equation}\label{dos}
     \text{DOS} (E)  = \frac{1}{2}\sum_{m', n'} A (m', n', E + \ri\eta).
    \end{equation}
On the graphs with  translational symmetry, a few states with increasing relative distance $(|m' - n'|)$ prove sufficient for converging results.

Substituting equation (\ref{ham}) in equation (\ref{gf}) generates the recursion relations between the Green's functions 
\begin{eqnarray}
&&(\omega - \epsilon_{m} -\epsilon_{n} - U\delta_{mn} - V_{m n}\delta_{m n\pm1})G_2(m,n,\omega) =   \delta_{m,m'}\delta_{n,n'} + \delta_{m,n'}\delta_{n,m'}   
-  t_{m-1, m}  G_2(m-1, n, \omega) \nonumber\\
 &&-  t_{m, m+1}  G_2(m+1, n, \omega) 
 - t_{n, n+1}  G_2(m, n+1, \omega) 
 - t_{n-1, n}  G_2(m, n-1, \omega)
\label{eq2.5}
\end{eqnarray}
for every site indices $m,n$ of the two particles. The Green's functions with constant $m+n$ are then grouped with a vector $\mathbf{V}_R$ ($R=m+n$) which forms a 1D chain and generates the recursion equation
   \begin{equation}\label{vrec}
    \mathbf{V}_{R} = \boldsymbol{\alpha}_R \mathbf{V}_{R-1} + \boldsymbol{\beta}_R \mathbf{V}_{R+1} + \mathbf{C}\,,
    \end{equation}
with $\bf C \neq \bf 0$ for $R = m' + n' = R'$ containing the initial state. The hopping matrices $\boldsymbol{\alpha}_R$, $\boldsymbol{\beta}_R$ connect Green's functions between nearest neighbour vectors with the hopping integrals. Solving this equation, any Green's function can be computed for a finite lattice. At the boundaries, equation (\ref{vrec}) becomes
   \begin{equation}\label{vlr}
     \mathbf{V}_{0} =  \boldsymbol{\beta}_0 \mathbf{V}_{1}   \mbox{     and     } \mathbf{V}_{L} = \boldsymbol{\alpha}_L \mathbf{V}_{L-1}\,,
    \end{equation}
 where $0$ and $L$ are the minimum and maximum indices for $R$. The recursion modifies to this general form
   \begin{equation}\label{vr}
    \mathbf{V}_{R} = \mathcal{A}_R \mathbf{V}_{R-1}   \mbox{   and   }  \mathbf{V}_{R} = \mathcal{B}_R \mathbf{V}_{R+1},   \mbox{             for  } R \neq R',
    \end{equation}
with
   \begin{equation}\label{rl}
 \mathcal{B}_{R} =  \left[1 -  \boldsymbol{\alpha}_R \mathcal{B}_{R-1}\right]^{-1} \boldsymbol{\beta}_R 
\mbox{    and     } 
  \mathcal{A}_{R} =  \left[1 -  \boldsymbol{\beta}_R \mathcal{A}_{R+1}\right]^{-1} \boldsymbol{\alpha}_R .
    \end{equation}
These $\mathcal{A}_{R}$ and $\mathcal{B}_{R}$ matrices can be computed recursively starting from equation (\ref{vlr}) before one reaches $R=R'$ from both sides of the chain with  
   \begin{eqnarray}\label{vr'}
  \mathbf{V}_{R'} = \left[1 - \boldsymbol{\alpha}_{R'}\mathcal{B}_{R'-1} - \boldsymbol{\beta}_{R'}\mathcal{A}_{R'+1}\right]^{-1} \mathbf{C}
    \end{eqnarray}
at $R= R'$. Once $\mathbf{V}_{R'}$ is found, all other  $\mathbf{V}_{R}$ are given by equation (\ref{vr}). The procedure accounts the full self-energy term which can be obtained from renormalized perturbation expansion \cite{econ} on any ordered or disordered lattice.  This method can also be used for the cases where sites have internal structures. 

In presence of external magnetic field, a gauge field appears in the Hofstadter Hamiltonian with the hopping terms which is known as Peierls substitution \cite{peie}. 

  \begin{eqnarray}\label{hof}
\mathcal{H}_{Hf} &=&  \sum_{m}  \left( t \re^{2\ri\piup \frac{p}{q} m_2} a_{m_1 +1}^\dagger a_{m_1} + t a_{m_2 +1}^\dagger a_{m_2} + h.c.\right) \nonumber\\
&+& \sum_m U a_{m}^\dagger a_{m}^\dagger a_m a_m + \sum_{<mn>} V_{|m-n|} a_{m}^\dagger a_{n}^\dagger a_n a_m\,.
 \end{eqnarray} 
For an ideal 2D lattice, the onsite energy terms can be scaled to zero with $q=\infty$ in the phase for zero magnetic field. The phase term contributes with the hopping on one of the axes of 2D lattice for non-zero magnetic field. The subscripts ${1,2}$ of the site index $m$ denotes coordinates of two axes of site~$m$. These terms are simulated in optical lattices with periodic modulation of lattice potentials \cite{dali}. The essence of the method depends on the classical Floquet theory where the momentum terms gain phases.

\section{Results}

The spectral density of bound pairs can be obtained from Green's functions computed using a recursion method or Chebyshev polynomials or exact diagonalization. The density of states had been computed before by Rausch et al. \cite{raus}, Halpap et al. \cite{half} and  others for one dimensional regular lattices. The density of states for two-dimensional ideal lattices had been computed before by Barelli et al. \cite{shep}. I use the method of recursion that has been formulated recently\cite{berc} by Berciu et al. for the computation of the spectral density of bound pairs at any interaction strength. In 1D, the recursive method provides converging results for the Green's functions with lattices of size in the order of $N=100$ sites.  The Green's functions are computed  for an 1D lattice with $N=500$ sites and $N=20\times21$ sites in 2D. For the Bethe lattice which resembles the structure of recursion, $L=8$ levels are considered. The value of hopping elements are fixed with $t=1$ while the interaction elements $U$ and $V$ are varied in the calculations.  

\begin{figure}[!t]
\centering
\includegraphics[width=0.7\textwidth]{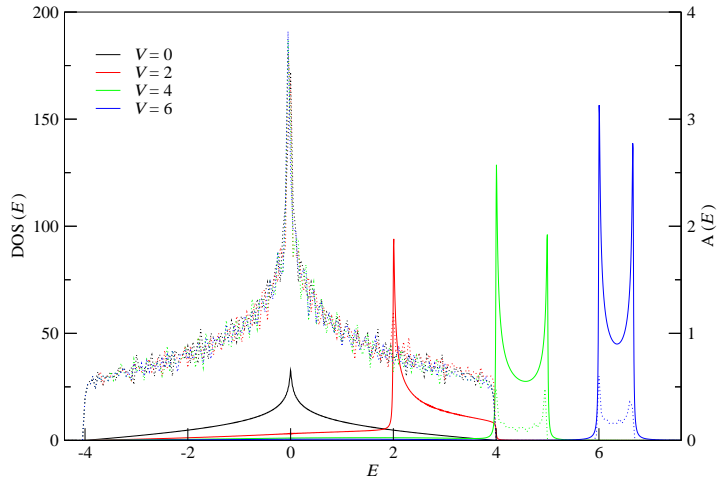}
\caption{(Colour online) Density of states of two interacting particles (dotted lines) and spectral density of bound pairs (straight lines) for a 1D lattice of $N=500$ computed from equation (\ref{spec}) and equation (\ref{dos}) to an arbitrary scale with $\eta=0.01$.}
\label{1Ds}
\end{figure}

The bound pair spectra in 1D have been recently studied by Rausch et. al for strong interactions. I obtained the results for weak interactions using the recursion which gives converging results at the band edges. The bound pairs within continuum can have different transport properties because of the scattering with the continuum states which I do not discuss in this article. The density of bound pairs at $T=0$ is maximum near $E = V$. The density of pairs at $V=0$ for noninteracting case is much smaller than for interacting $V\neq 0$ cases. The bound state splits from the continuum at $V=W/2$ where $W$ is the bandwidth of the two-particle continuum spectra.  The bound state shows a sharp peak near $E = V$ for weak interaction strengths ($V<W/2$). For strong interaction cases, the density reflects the cosine dispersion curve of bound states \cite{petr} near $E=V$. The total density of states is similar to the density of states that is obtained analytically for the noninteracting single particles in 2D lattices \cite{econ} except the satellite peaks which are clearly observed for the strong interaction cases.

\begin{figure}[!t]
\begin{subfigure}{0.326\textwidth}
\includegraphics[width = 1.13\textwidth]{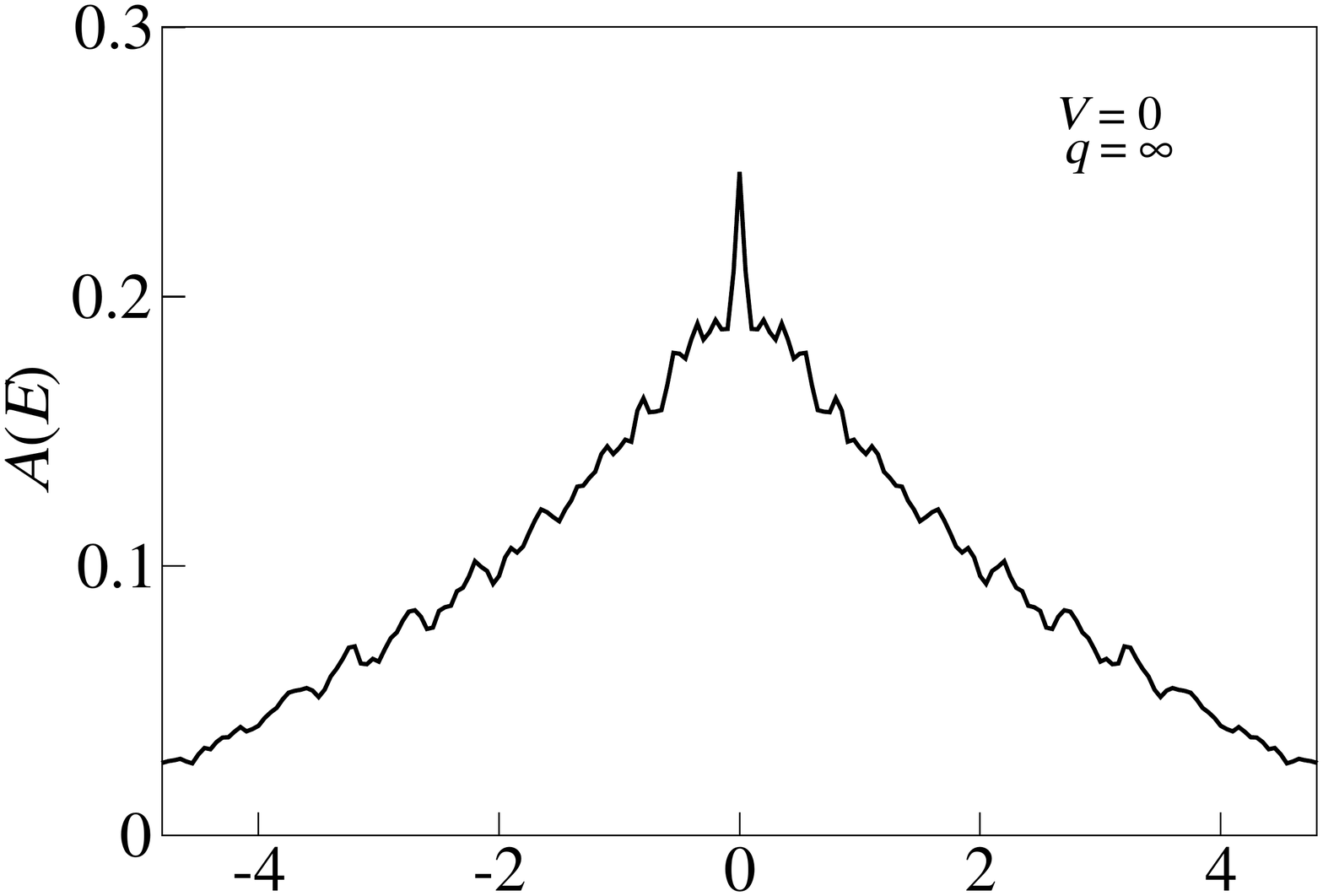}
\end{subfigure}
\begin{subfigure}{0.326\textwidth}
\includegraphics[width = 1.13\textwidth]{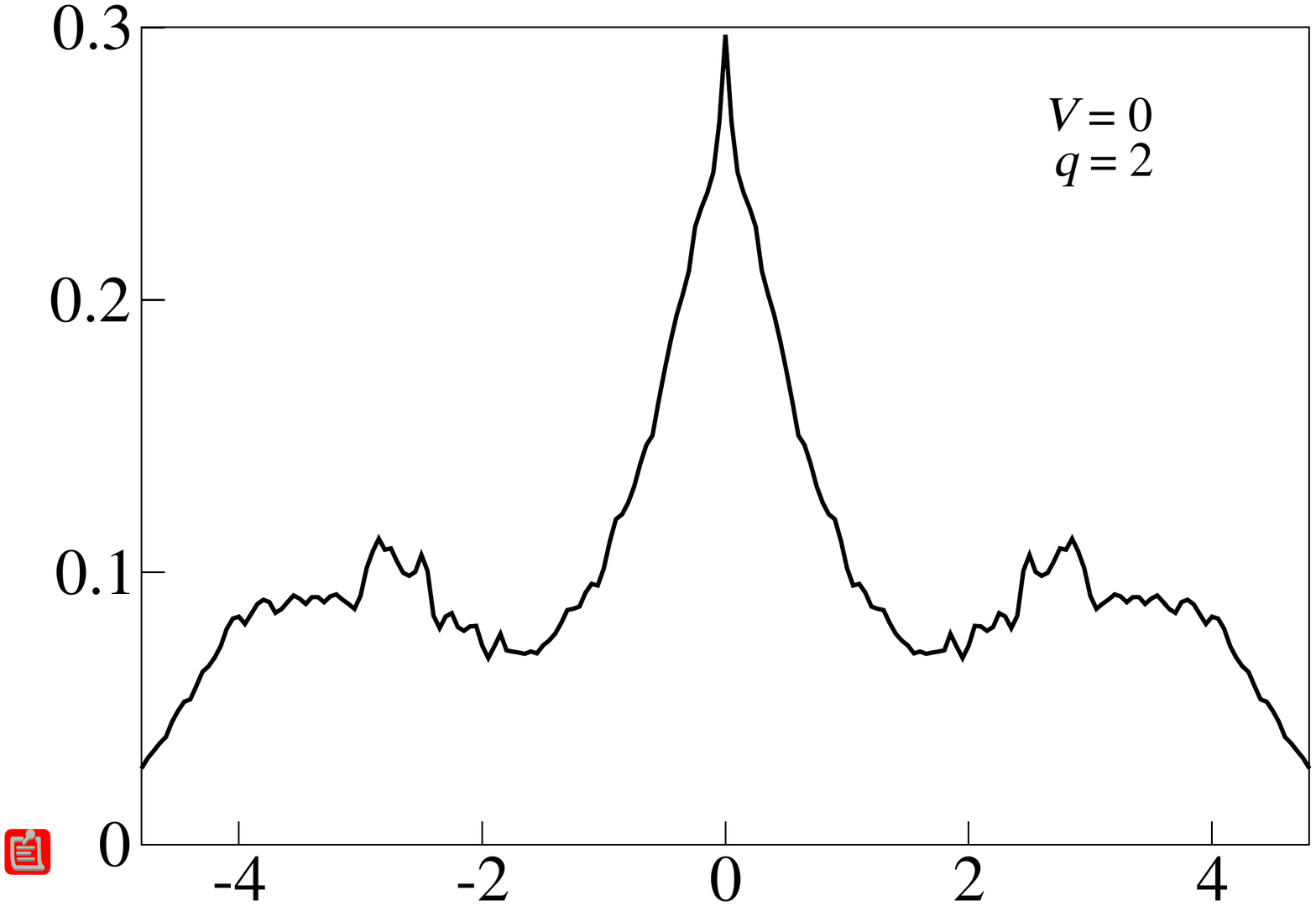}
\end{subfigure}
\begin{subfigure}{0.326\textwidth}
\includegraphics[width = 1.13\textwidth]{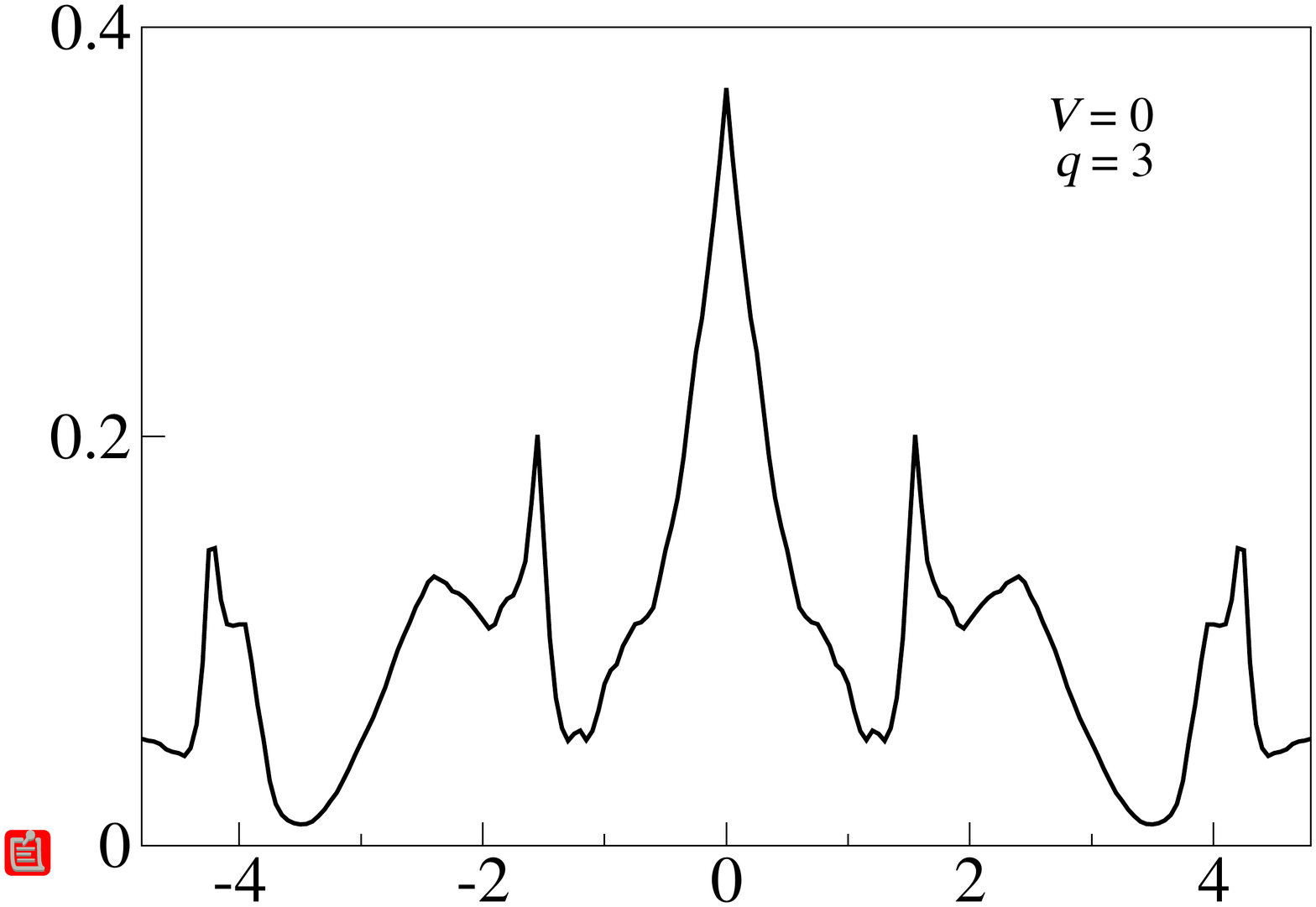}
\end{subfigure}
\vspace{-0.1cm}
\begin{subfigure}{0.326\textwidth}
\includegraphics[width = 1.13\textwidth]{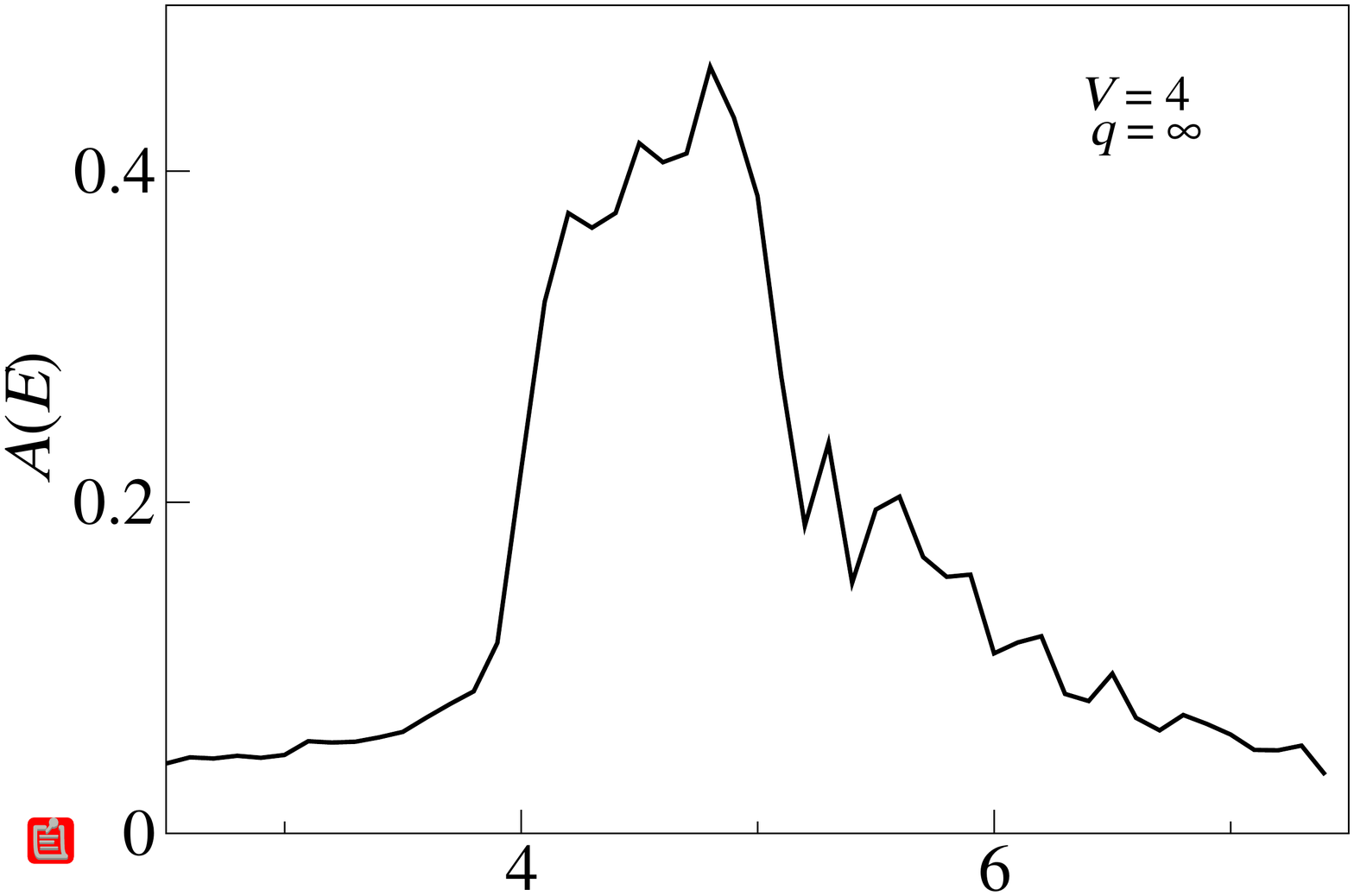}
\end{subfigure}
\begin{subfigure}{0.326\textwidth}
\includegraphics[width = 1.13\textwidth]{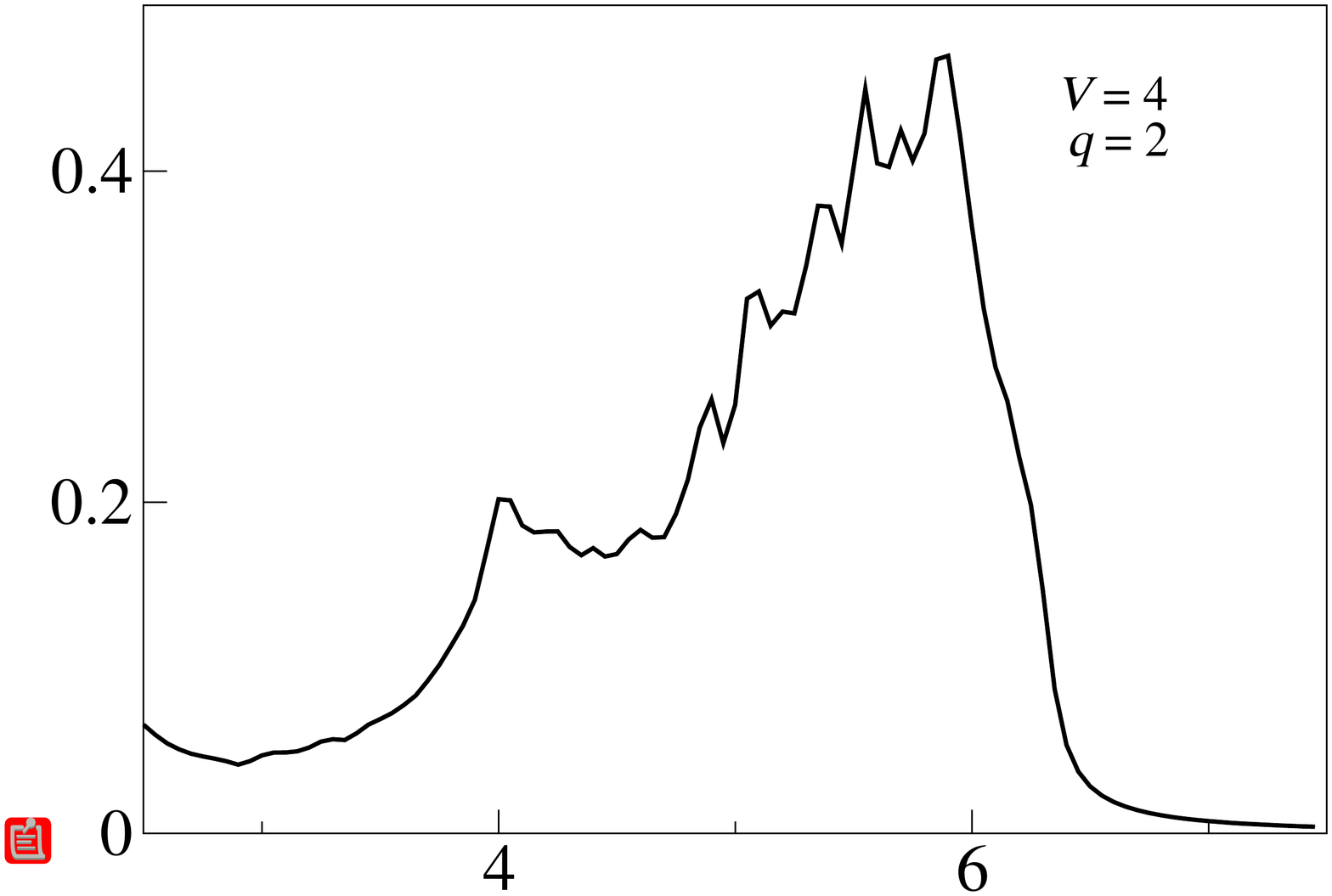}
\end{subfigure}
\begin{subfigure}{0.326\textwidth}
\includegraphics[width = 1.13\textwidth]{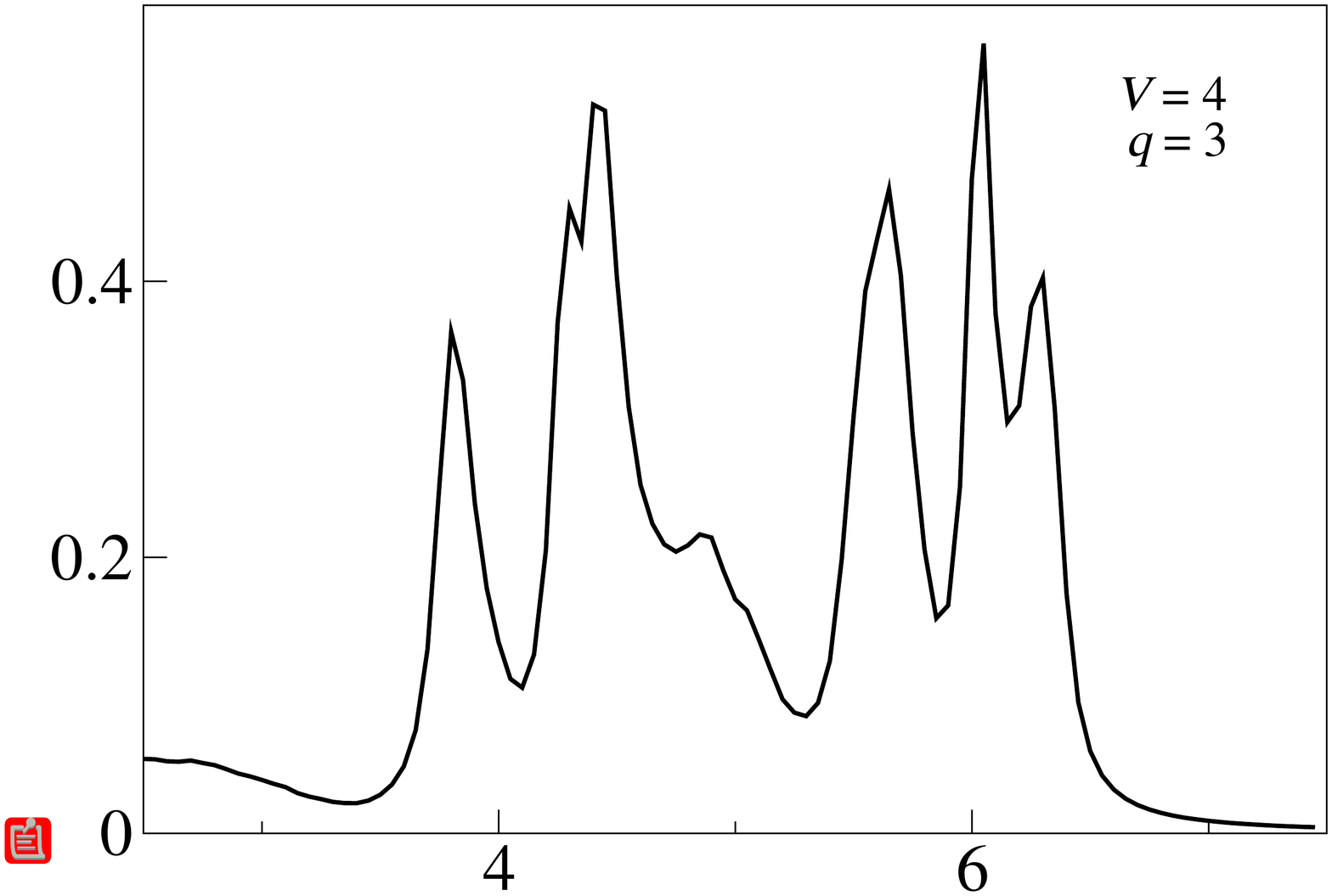}
\end{subfigure}
\begin{subfigure}{0.326\textwidth}
\includegraphics[width = 1.13\textwidth]{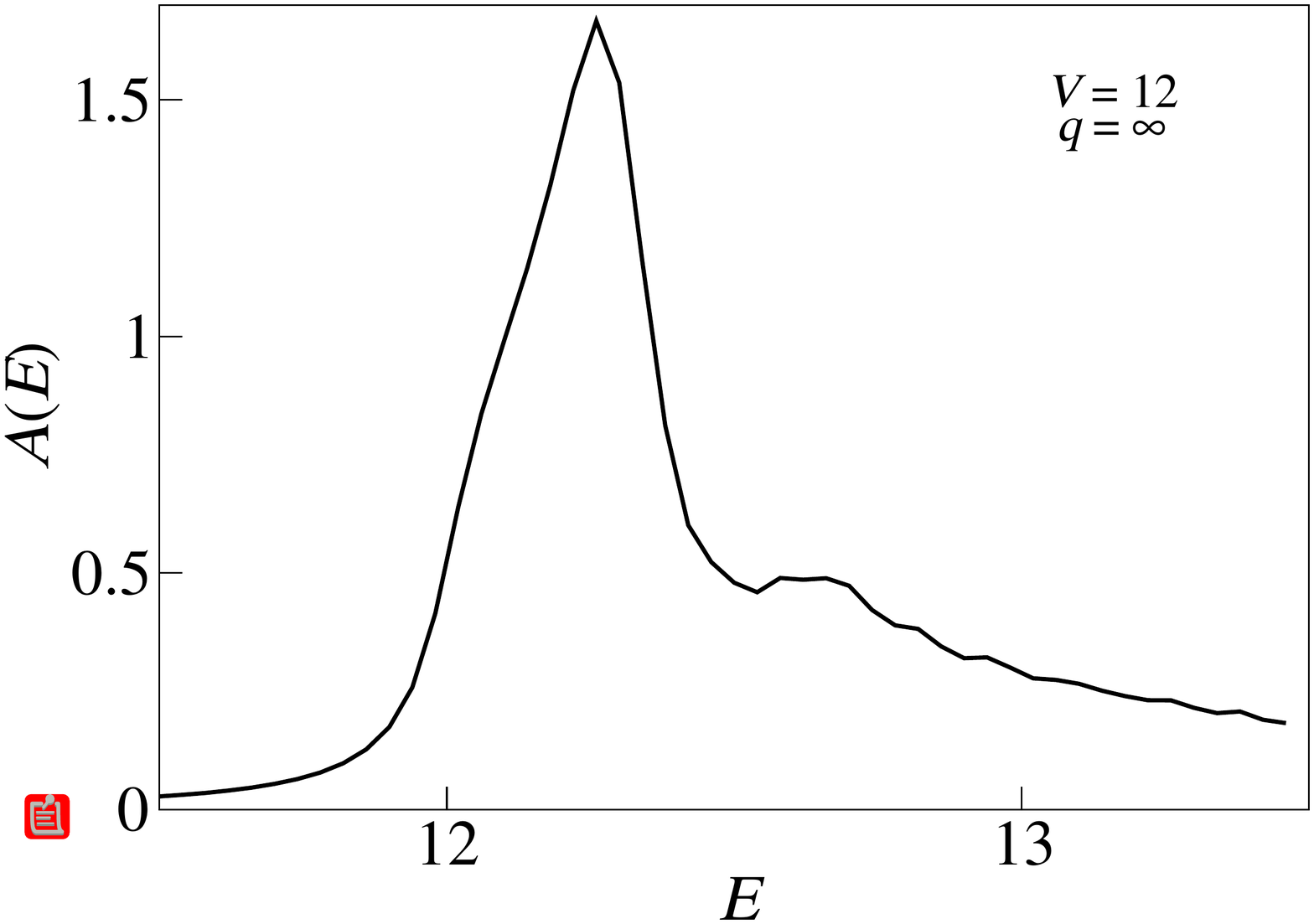}
\end{subfigure}
\begin{subfigure}{0.326\textwidth}
\includegraphics[width = 1.13\textwidth]{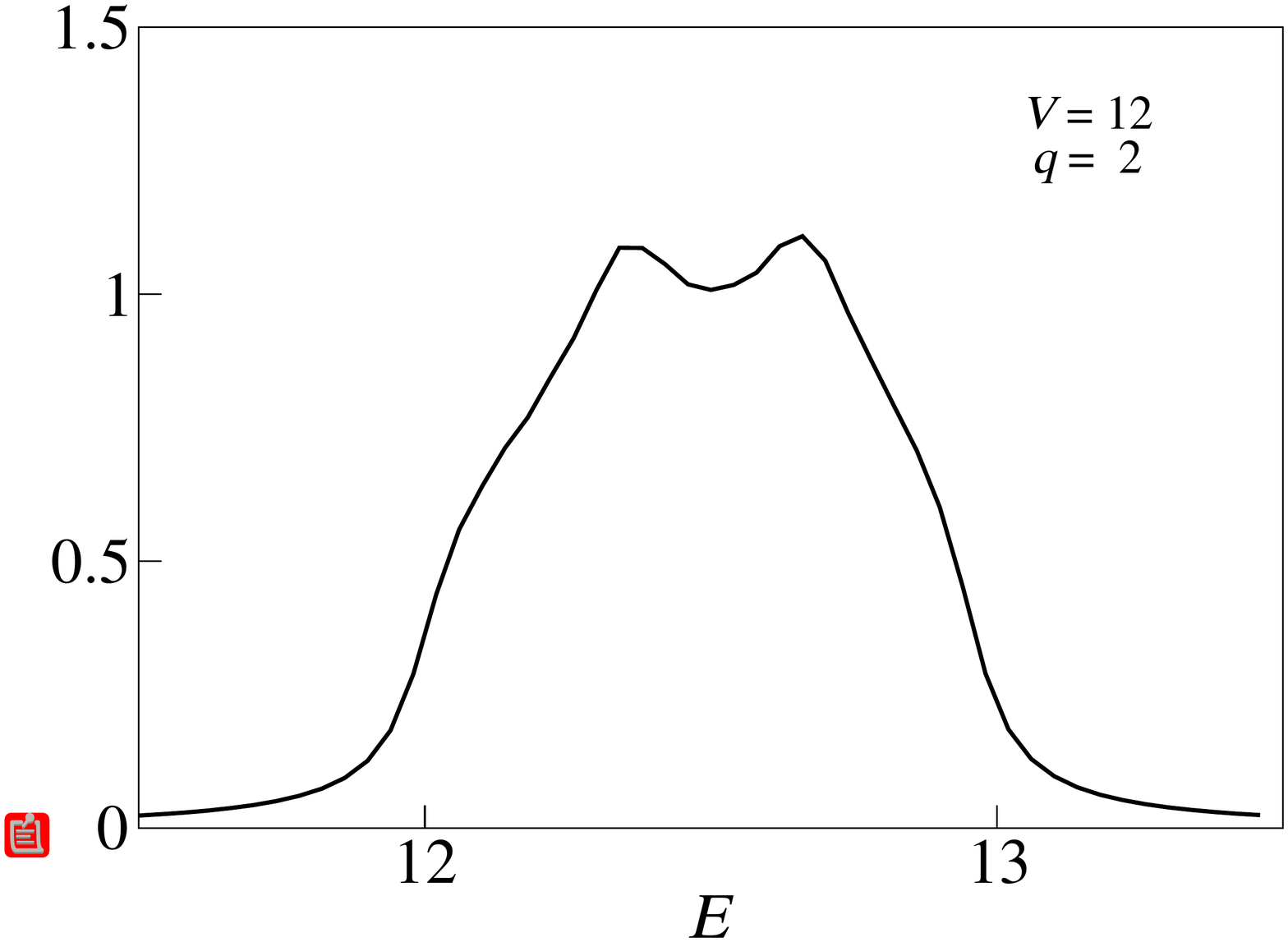}
\end{subfigure}
\begin{subfigure}{0.326\textwidth}
\includegraphics[width = 1.13\textwidth]{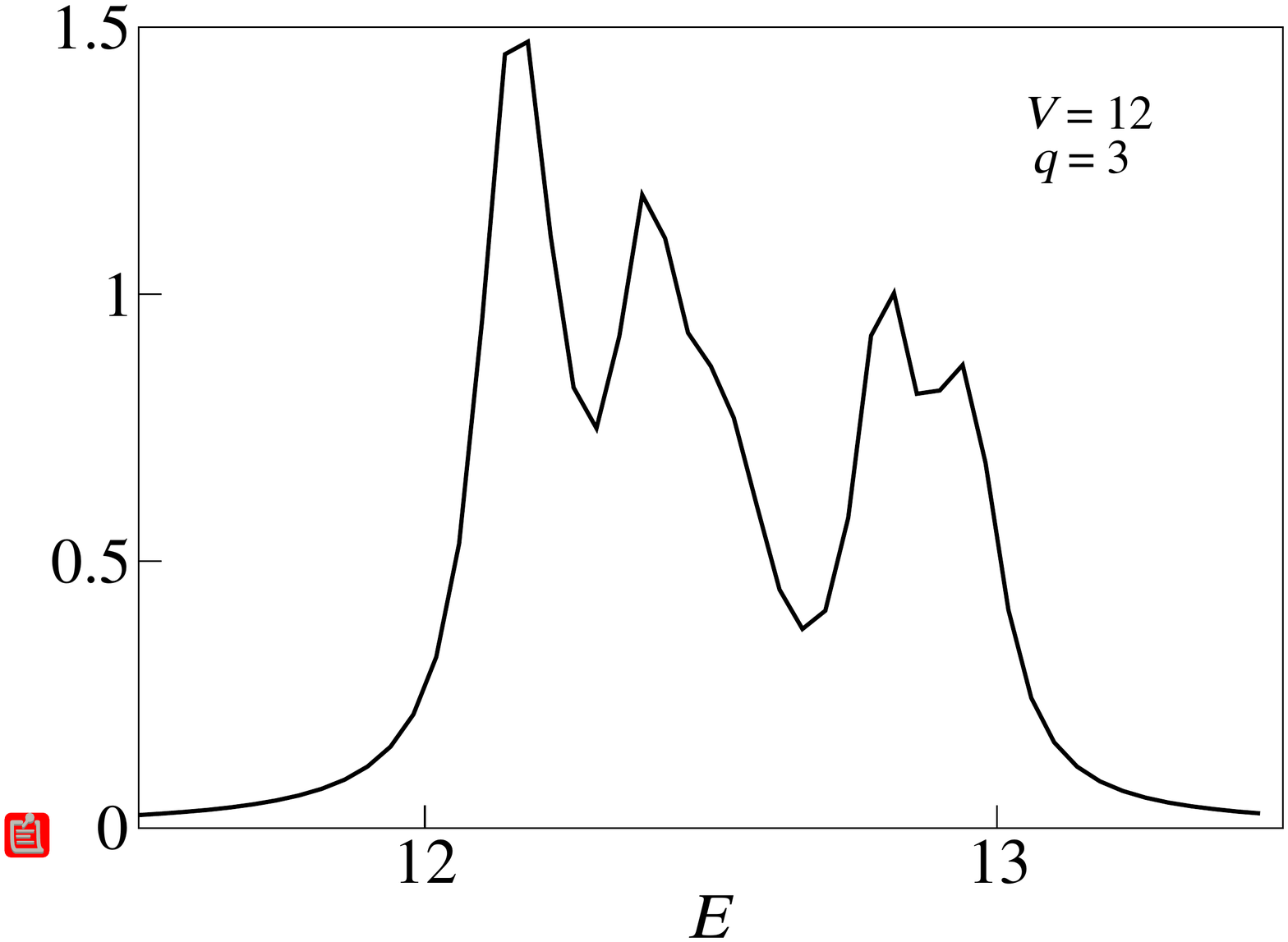}
\end{subfigure}
\caption{Spectral density of bound pairs on a 2D lattice of $N=20\times21$ with $p=1$ computed from equation (\ref{spec}) for the Hofstadter Hamiltonian in equation (\ref{hof}) with $U=\infty, t_1 = 1, t_2 = 1$ and $\eta= 0.06$.}
\label{2Ds}
\end{figure}

The bound pair density shows a similar structure of one bound state in 2D lattices in presence of interaction. The density profiles are very similar to that of 1D lattice for $V=0$. For weak interaction strengths, the pair density reaches the maximum within a range of the bandwidth rather than a sharp peak and decays to the band edge afterwords. For strong interactions, the pair density shows a sharp maximum at $V>W/2$. In presence of external gauge fields, splitting in the spectra can be observed from figure \ref{2Ds}.  The details of the spectra then depend on the flux per plaquette $\frac{p}{q}$. At half flux per plaquette for the non-interacting cases, the pair density is reminiscent of Kondo profile. At strong interactions with half flux per plaquette, the density is regular dome-shaped while at non-half fluxes, the density is highly irregular. At weak interaction, the maximum density is shifted to higher $E$. These density profiles will be reflected symmetrically about $E=0$ with changes of sign in the interaction term $V \rightarrow -V$. 

\begin{figure}[!t]
\begin{subfigure}{0.485\textwidth}
\includegraphics[width = 1.0\textwidth]{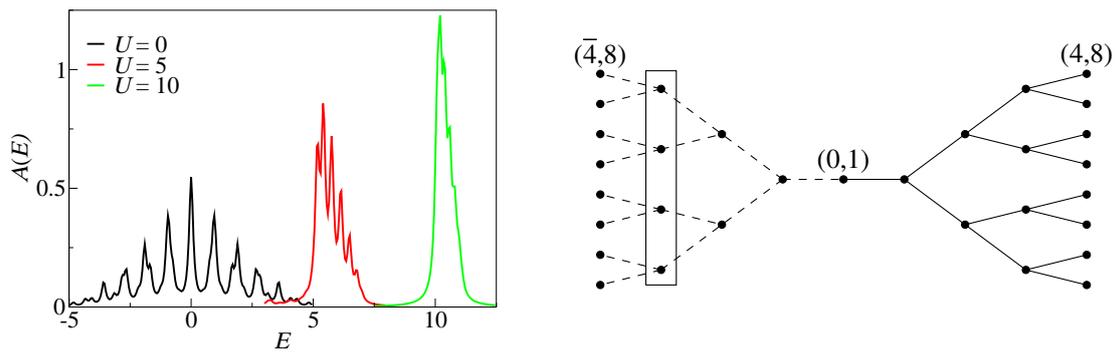}
\end{subfigure}
\begin{subfigure}{0.485\textwidth}
\begin{tikzpicture}[scale=0.8]
\draw [fill] (0,0) circle [radius=0.06];
\draw [fill] (0,0.5) circle [radius=0.06];
\draw [fill] (0,1) circle [radius=0.06];
\draw [fill] (0,1.5) circle [radius=0.06];
\draw [fill] (0,2) circle [radius=0.06];
\draw [fill] (0,2.5) circle [radius=0.06];
\draw [fill] (0,3) circle [radius=0.06];
\draw [fill] (0,3.5) circle [radius=0.06];
\draw [fill] (1,0.25) circle [radius=0.06];
\draw [fill] (1,1.25) circle [radius=0.06];
\draw [fill] (1,2.25) circle [radius=0.06];
\draw [fill] (1,3.25) circle [radius=0.06];
\draw [fill] (2,1) circle [radius=0.06];
\draw [fill] (2,2.5) circle [radius=0.06];
\draw [fill] (3,1.75) circle [radius=0.06];
\draw [fill] (4,1.75) circle [radius=0.06];
\draw [fill] (5,1.75) circle [radius=0.06];
\draw [fill] (6,1) circle [radius=0.06];
\draw [fill] (6,2.5) circle [radius=0.06];
\draw [fill] (7,0.25) circle [radius=0.06];
\draw [fill] (7,1.25) circle [radius=0.06];
\draw [fill] (7,2.25) circle [radius=0.06];
\draw [fill] (7,3.25) circle [radius=0.06];
\draw [fill] (8,0) circle [radius=0.06];
\draw [fill] (8,0.5) circle [radius=0.06];
\draw [fill] (8,1) circle [radius=0.06];
\draw [fill] (8,1.5) circle [radius=0.06];
\draw [fill] (8,2) circle [radius=0.06];
\draw [fill] (8,2.5) circle [radius=0.06];
\draw [fill] (8,3) circle [radius=0.06];
\draw [fill] (8,3.5) circle [radius=0.06];
\draw (4,1.75) --(5,1.75);
\draw (6,1)  --(5,1.75);
\draw (6,2.5)  --(5,1.75);
\draw (6,1)  --(7,0.25);
\draw (6,1)  --(7,1.25);
\draw (6,2.5)  --(7,2.25);
\draw (6,2.5)  --(7,3.25);
\draw (8,0)  --(7,0.25);
\draw (8,1)  --(7,1.25);
\draw (8,2)  --(7,2.25) ;
\draw (8,3)  --(7,3.25) ;
\draw (8,0.5)  --(7,0.25);
\draw (8,1.5)  --(7,1.25);
\draw (8,2.5)  --(7,2.25);
\draw (8,3.5)  --(7,3.25);
\draw [dashed] (3,1.75) --(4,1.75);
\draw [dashed] (3,1.75) --(2,2.5);
\draw [dashed] (3,1.75) --(2,1);
\draw [dashed] (1,2.25) --(2,2.5);
\draw [dashed] (1,0.25) --(2,1);
\draw [dashed] (1,3.25) --(2,2.5);
\draw [dashed] (1,1.25) --(2,1);
\draw [dashed] (1,2.25) --(0,2);
\draw [dashed] (1,0.25) --(0,0);
\draw [dashed] (1,3.25) --(0,3);
\draw [dashed] (1,1.25) --(0,1);
\draw [dashed] (1,2.25) --(0,2.5);
\draw [dashed] (1,0.25) --(0,0.5);
\draw [dashed] (1,3.25) --(0,3.5);
\draw [dashed] (1,1.25) --(0,1.5);
\draw (0.75,0) -- (1.25,0) -- (1.25,3.55) -- (0.75,3.55) -- (0.75,0);
\node at (4,2.05) {(0,1)};
\node at (8,3.8) {(4,8)};
\node at (0,3.85) {($\overline{4}$,8)};
\end{tikzpicture}
\end{subfigure}
\caption{(Colour online) Spectral density of bound pairs at various onsite interaction strengths on a Bethe lattice of $L=8$ with $\eta = 0.1$. The Bethe lattice is depicted on the right.}
\label{trees}
\end{figure}

The bound pair state shows non-trivial features on Bethe lattices. At zero onsite interaction $U=0$, the spectra show discontinuity and multiple sharp peaks. At strong interactions, these peaks merge within a single envelope. The pair density in figure \ref{trees} is obtained for two bosons, and the interactions between particles at different nodes are neglected. The initial states are prepared at the root layer $L=0$ with two bosons on the node ($L=0$, $n=1$). For two particles, the recursion can be performed with $R=L_1 + L_2$ adding layer indices for the two particles on the Bethe lattice.  For single particles, the recursion vectors map directly to the branches of each layer divided as $\overline{L}$ and $L$ as depicted in figure \ref{trees}. The Green's functions can be then efficiently computed for any initial superposition of states on these vectors, e.g., the rectangular box on the graph will be the vector $\mathbf{V}_1$ in the recursion. The discontinuities of the spectra for weak interactions within small energy width suggest tunable control of bound pair transport on these graphs.

\section{Conclusion}
In this article the doublon spectra of regular 1D, 2D and Bethe lattices have been computed from real space Green's functions. In two dimensions, the effects of external magnetic fields on the spectra have been computed for weak and strong nearest neighbour interactions between two hardcore bosons. For weak interactions, the gauge fields enhance the density at few bandwidths. In Bethe lattices, the spectra show discontinuities for weak interactions within the band.  This work points to energy windows where maximum bound pairs can be found in 1D, 2D and Bethe lattice systems with the change of interaction strengths between particles. The work might be helpful for understanding the many-particle states where bound pairs can lead to the difference in responses, e.g., conductivity and spectroscopy of the lattice systems. The scattering and dissipation of the bound pairs within continuum should also be analyzed. This work computes the spectral profiles which provide insights into the mechanism of interacting particles when two particle bound pair densities are much higher than the bound states with a larger number of particles. Many-particle bound states with  a larger number of particles at different energies should be studied further which can also be important for the systems with strong interactions.

\subsection*{Acknowledgements}
The author acknowledges funding support from NSERC, Canada.

\ukrainianpart

\title{Спектральна густина  взаємодіючих пар у низьковимірних скінчених гратках%
}
\author{T. Чаттарай} 
\address{
Унiверситет Британської Колумбiї (UBC), Ванкувер, Канада
}
\makeukrtitle 
\begin{abstract}
Обчислено спектральну густину зв'язаних пар в ідеальних одно-,  двовимірних і Бете гратках  для слабких та сильних взаємодій. Обчислення здійснено з допомогою   функцій Гріна ефективним рекурентним методом в реальному просторі. Для діапазону сил взаємодії, в межах якого зв’язані стани є переважно одинарними парами, спектральні профілі скеровані на таку пропускну здатність енергії, де зв’язані пари можуть бути максимізовані.

\keywords квазічастинки та колективні збудження, переходи метал-ізолятор, дублони, зв'язані стани
\end{abstract}


\begin{thebibliography}{50}
\bibitem{hubb} Hubbard J., Proc. R. Soc. London, Ser. A, 1963, \textbf{276}, 238, \doi{10.1098/rspa.1963.0204}.

\bibitem{wink} Winkler K., Thalhammer G., Lang F., Grimm R., Hecker Denschlag J., Daley A.J., Kantian A., B\"uchler H.P., Zoller P., Nature 2006, \textbf{441}, 853, \doi{10.1038/nature04918}.

\bibitem{sawa} Sawatzky G.A., Phys. Rev. Lett. 1977, \textbf{39}, 504, \doi{10.1103/PhysRevLett.39.504}. 
\bibitem{lens} Sawatzky G.A., Lenselink A.L., Phys. Rev. B. 1980, \textbf{21}, 1790, \doi{10.1103/PhysRevB.21.1790}. 

\bibitem{cini} Cini M., J. Phys.: Condens. Matter, 1989, \textbf{1}, SB55, \doi{10.1088/0953-8984/1/SB/011}.

\bibitem{stro} Strohmaier N., Greif D., J\"ordens R., Tarruell L., Moritz H., Esslinger T., Sensarma R., Pekker D., Altman E., Demler E., Phys. Rev. Lett., 2010, \textbf{104}, 080401, \doi{10.1103/PhysRevLett.104.080401}.

\bibitem{prei}  Preiss P.M., Ma R., Tai E.M., Lukin A., Rispoli M., Zupancic P., Lahini Y., Islam R., Greiner M., Science, 2015, \textbf{347}, 1229, \doi{10.1126/science.1260364}.

\bibitem{fuku} Fukuhara T., Schau\ss~P., Endres M., Hild S., Cheneau M., Bloch I., Gross C., Nature, 2013, \textbf{502}, 72, \\\doi{10.1038/nature12541}.

\bibitem{klei} Klein A., Krejs F., Phys. Rev.,  1976, \textbf{13}, 3282. 
\bibitem{mill} Miller P.D., Scott A.C., Carr J., Eilbeck J.C., Phys. Scr., 1991, \textbf{44}, 509, \doi{10.1088/0031-8949/44/6/002}.

\bibitem{qin} Qin X., Ke Y., Guan X., Li Z., Andrei N., Lee C., Phys. Rev. A, 2014, \textbf{90}, 062301, \\\doi{10.1103/PhysRevA.90.062301}. 

\bibitem{mcwe} McWeeny R., Rev. Mod. Phys., 1960, \textbf{32}, 335.  

\bibitem{chil} Childs A.M., Goldstone J., Phys. Rev. A, 2004, \textbf{70}, 022314, \doi{1050-2947/2004/70(2)/022314(11)}.
\bibitem{child} Childs A.M., Gosset D., Web Z., Science, 2013, \textbf{339}, 791, \doi{10.1126/science.1229957}.

\bibitem{perr} Peruzzo A., Lobino M., Matthews J.C.F., Matsuda N., Politi A., Poulios K., Zhou X.-Q., Lahini Y., Ismail~N., W\"orhoff K., Bromberg Y., Silberberg Y., Thompson M.G., OBrien J.L., Science, 2010, \textbf{329}, 1500, \doi{10.1126/science.1193515}.

\bibitem{bard} Bardeen J., Cooper L.N.,  Schrieffer J.R., Phys. Rev., 1957, \textbf{108}, 1175, \doi{10.1103/PhysRev.108.1175}. 

\bibitem{micn} Micnas R., Ranninger J., Robaskiewicz S., Rev. Mod. Phys., 1990, \textbf{62} 113, \doi{10.1103/RevModPhys.62.113}.

\bibitem{maju} Majumdar S., Solid State Commun. 1988, \textbf{66}, 427, \doi{10.1016/0038-1098(88)90870-8}. 

\bibitem{hofs}  Hofstadter D.R., Phys. Rev. B 1976, \textbf{14}, 2239, \doi{10.1103/PhysRevB.14.2239}.

\bibitem{shep} Barelli A., Bellissard J., Jacquod P., Shepelyansky D.L., Phys. Rev. B, 1997, \textbf{55}, 9524, \\\doi{10.1103/PhysRevB.55.9524}.

\bibitem{lin} Lin Y.-J., Compton R.L., Jim\'{e}nez-Garc\'{i}a K., Porto J.V., Spielman I.B., Nature, 2009, \textbf{462}, 628, \\\doi{10.1038/nature08609}. 

\bibitem{dali} Dalibard J., Gerbier F., Juzeli\~{u}nas G., \"Ohberg P., Rev. Mod. Phys., 2011, \textbf{83}, 1523, \\\doi{10.1103/RevModPhys.83.1523}.

\bibitem{stru} Struck J., \"Olschl\"ager C., Weinberg M., Hauke P., Simonet J., Eckardt A.,  Lewenstein M., Sengstock K., Windpassinger P., Phys. Rev. Lett., 2012, \textbf{108}, 225304, \doi{10.1103/PhysRevLett.108.225304}. 

\bibitem{petr} Valiente M., Petrosyan D., J. Phys. B: Mol. Opt. Phys., 2008, \textbf{41}, 161002, \doi{10.1088/0953-4075/41/16/161002}. 

\bibitem{recu} Pettifor D.G., Weaire D.L., The Recursion Method and Its Applications, Springer-Verlag, London, 1984, \doi{10.1007/978-3-642-82444-9}.

\bibitem{lanc} Lanczos C., J. Res. Nat. Bur. Stand., 1950, \textbf{45}, 255, \doi{10.6028/jres.045.026}. 

\bibitem{scho} Schollow\"ock U., Ann. Phys., 2011, \textbf{326}, 96, \doi{10.1016/j.aop.2010.09.012}. 

\bibitem{berc} Berciu M., Cook A.M., EPL, 2010, \textbf{92}, 40003, \doi{10.1209/0295-5075/92/40003}

\bibitem{tirt} Chattaraj T., Condens. Matter, 2018, \textbf{3}, 38, \doi{10.3390/condmat3040038}. 

\bibitem{econ} Economou E.N., Green's Functions in Quantum Physics, Springer-Verlag, London, 2006. 

\bibitem{peie} Peierls R., Z. Phys., 1933, \textbf{80}, 763.

\bibitem{raus} Rausch R., Potthoff M., New J. Phys., 2016, \textbf{18}, 023033, \doi{10.1088/1367-2630/18/2/023033}.

\bibitem{half} Halfpap O., Zharekeshev I.K., MacKinnon A., Kramer B., Ann. Phys.,  1998, \textbf{7}, 503, \\\doi{10.1002/(SICI)1521-3889(199811)7:5/6\%3C503::AID-ANDP503\%3E3.0.CO;2-L}. 



\end{thebibliography}
\end{document}